\documentclass[10pt,journal,compsoc]{IEEEtran}

\usepackage{amsmath}
\usepackage{enumitem}
\usepackage{ragged2e}
\usepackage{makecell}
\usepackage{multirow}
\usepackage[hyphens]{url}
\usepackage{hyperref}
\usepackage{color, colortbl}
\usepackage{array}
\usepackage[caption=false,font=footnotesize,labelfont=sf,textfont=sf]{subfig}

\ifCLASSOPTIONcompsoc
  \usepackage[nocompress]{cite}
\else
  \usepackage{cite}
\fi
\ifCLASSINFOpdf
  \usepackage[pdftex]{graphicx}
\else
  \usepackage[dvips]{graphicx}
\fi

\newcommand{\rev}[1]{{\color{black}#1}}
\title{An In-Module Disturbance Barrier for Mitigating Write Disturbance in Phase-Change Memory} %ACM

\author{Hyokeun~Lee,~\IEEEmembership{Member,~IEEE}
        Seungyong~Lee,
        Byeongki~Song,
        Moonsoo~Kim,
        Seokbo~Shim,
        Hyuk-Jae~Lee,~\IEEEmembership{Member,~IEEE}
        and~Hyun~Kim,~\IEEEmembership{Member,~IEEE}%

\IEEEcompsocitemizethanks{%
\IEEEcompsocthanksitem H.~Lee, S.~Lee, H.~J.~Lee are with Inter-university of Semiconductor Research Center, Department of Electrical and Computer Engineering, Seoul National University, Seoul, South Korea. %\proect\\
E-mail: \{hklee, sylee, hyuk\_jae\_lee\}@capp.snu.ac.kr
\IEEEcompsocthanksitem S.~Shim is with SK Hynix Inc., Icheon, Gyeonggi-do, South Korea. %\protect\\
E-mail: seokbo.shim@sk.com
\IEEEcompsocthanksitem M.~Kim and B.~Song are with Samsung Inc., Hwasung, Gyeonggi-do, South Korea. %\protect\\ 
E-mail: ms213.kim@samsung.com, bksong@capp.snu.ac.kr
\IEEEcompsocthanksitem H. Kim is with the Department of Electrical and Information Engineering and the Research Center for Electrical and Information Technology, Seoul National University of Science and Technology, Seoul, South Korea. %\protect\\
(Corresponding author: Hyun Kim.) 
Email: hyunkim@seoultech.ac.kr
}%
\thanks{Manuscript received 04 Nov. 2021; accepted 23 July 2022(DOI 10.1109/TC.2022.3197071).
}
}

\begin{document}

\bstctlcite{IEEEexample:BSTcontrol}

\pagenumbering{arabic}

%%%%%% -- PAPER CONTENT STARTS-- %%%%%%%%
\IEEEtitleabstractindextext{%
\begin{abstract}
\justifying 
\rev{Write disturbance error (WDE) appears as a serious reliability problem preventing phase-change memory (PCM) from general commercialization, and therefore several studies have been proposed to mitigate WDEs. Verify-and-correction (VnC) eliminates WDEs by always verifying the data correctness on neighbors after programming, but incurs significant performance overhead. Encoding-based schemes mitigate WDEs by reducing the number of WDE-vulnerable data patterns; however, mitigation performance notably fluctuates with applications. Moreover, encoding-based schemes still rely on VnC-based schemes. Cache-based schemes lower WDEs by storing data in a write cache, but it requires several megabytes of SRAM to significantly mitigate WDEs. Despite the efforts of previous studies, these methods incur either significant performance or area overhead. Therefore, a new approach, which does not rely on VnC-based schemes or application data patterns, is highly necessary. Furthermore, the new approach should be transparent to processors (i.e., in-module), because the characteristic of WDEs is determined by manufacturers of PCM products. In this paper, we present an in-module disturbance barrier (IMDB) that mitigates WDEs on demand. IMDB includes a two-level hierarchy comprising two SRAM-based tables, whose entries are managed with a dedicated replacement policy that sufficiently utilizes the characteristics of WDEs. The naive implementation of the replacement policy requires hundreds of read ports on SRAM, which is infeasible in real hardware; hence, an approximate comparator is also designed. We also conduct a rigorous exploration of architecture parameters to obtain a cost-effective design. The proposed method significantly reduces WDEs without noticeable speed degradation or additional energy consumption compared to previous methods.}

\end{abstract}

}

\maketitle
% Stuffs for squeezing paper
%\linespread{0.9} % For squeezing without noticing
%TODO: PCMCsim reference 추가하기!!!
\section{Introduction} 
\label{S1}

Phase-change memory (PCM) is gaining attention as the next-generation non-volatile memory (NVM), owing to its non-volatility, low latency, and scalability \cite{PCMAlter}. In recent years, software-defined memory has been announced to utilize NVM as high-speed storage or extended memory interchangeably \cite{IMDTGuide}. In particular, in-memory databases require data to remain in memory and be accessible with low latency; hence, a high-performance database can be developed by employing PCM as a non-volatile main memory \cite{NVHeaps, ATOM, REDU}. Moreover, products of PCM have been tested in various environments for evaluating performance and exploring their suitable applications \cite{DCPMMandHPSciApp, SysEvalDCPMM}. Therefore, leveraging and enhancing PCM-related technology is crucial to attaining low-latency and large-scale memory systems in the future. 

Even though PCM has attractive characteristics, it is not ready to be popularized in the consumer market, because several reliability issues still exist in PCM \cite{MultiwayWL, WSHR, CEnT, SIWC, RMWMerge, TurboRead}. In particular, write disturbance error (WDE) is one of the major problems, which delays its widespread commercialization. WDE is an interference problem on adjacent cells similar to row-hammer in DRAM \cite{DRAMDisturb}. This problem must be addressed as the highest priority because it would be exacerbated as process technology shrinks \cite{ADAM}. Additionally, in-memory database directly store data in NVM by utilizing cache-line flushes \cite{ATOM, REDU}. This kind of application would incur frequent write operations, making cells vulnerable to WDEs.

Previously, various approaches have been reported to mitigate WDEs in PCM devices \cite{SDPCM, DIN, ADAM, MinWD, SIWC, CEnT,DCPCM}. Approaches based on verification-and-correction (VnC) are able to eliminate all WDEs \cite{VnCinSTT, SDPCM}. However, VnC incurs \rev{additional} read operations for checking the existence of errors, degrading the performance significantly. Encoding-based schemes \cite{DIN, MinWD, ADAM, WLC, CEnT, ADAPT} reduce the number of WDE-vulnerable data patterns with little reliance on VnC, but the mitigation performance of these approaches varies considerably with data patterns in applications. Studies \cite{SIWC, ModelPCM2} have reported that WDEs may occur when a cell experiences more than a specific number of RESET pulses from its neighbors, which is more realistic than a random WDE model. Although the study in\cite{SIWC} has presented the manufacturing metric that incurs WDEs, their approach leverages a write-cache to reduce the write traffic without considering such a realistic WDE model. Furthermore, a large capacity of SRAM is required for mitigating WDEs notably. For above reasons, given that previous approaches are entirely decoupled from this realistic model, \rev{new approaches that manage aggressors, which are actively programmed cells that likely incur WDEs on neighboring cells, are necessary with negligible performance overhead in PCM modules.}

To satisfy these \rev{requirements}, this paper proposes an in-module write disturbance barrier (IMDB) that utilizes a realistic WDE model and restores vulnerable data on demand. Because the realistic WDE model shows that WDEs occur with a specific number of neighboring writes, the proposed method records the number of RESETs in a table. Using the recorded information, most of the WDE-vulnerable data can be rewritten before the occurrences of WDEs, and only addresses need to be managed in the data structure to reduce the burden on the supercapacitors upon system failure. For further error mitigation, a tiny data cache, referred to as a \textit{barrier buffer}, is introduced to store highly aggressive address information. Meanwhile, the replacement policy may expand the number of read ports on SRAM, involving a considerable overhead. This is because the policy merely regards the entry holding a smaller number of 1-to-0 flips as an eviction candidate. Therefore, an approximate lowest number estimator (\textit{AppLE}), which probabilistically counts the numbers based on the sampling method, is proposed to accommodate the use of a dual-port SRAM (DPSRAM) without speed degradation. Experimental results indicate that our approach reduces WDEs compared to previous studies, with negligible overhead. In conclusion, the contributions of this study can be summarized as follows:

\begin{itemize}
\item The first on-demand WDE mitigation method is proposed. Based on a more practical WDE trigger model, the proposed method leverages a two-level SRAM and restores vulnerable cells on demand.
\item This paper introduces a novel prior-knowledge-offering method, because the replacement policy may contradict the locality of applications. 
\item The replacement policy requires hundreds of ports on an SRAM in a naive approach. This paper designs probabilistic hardware, AppLE, to allow the use of a DPSRAM for enhancing the feasibility.
\item Several design parameters are required in the proposed method; hence, rigorous sensitivity analyses are conducted to acquire the cost-effective design.
\end{itemize}

\section{Background and Motivation} \label{S2}

\subsection{Introduction to Phase-Change Memory} \label{S2.1}

\begin{figure} [t]
	\centering
	\includegraphics[scale=0.85]{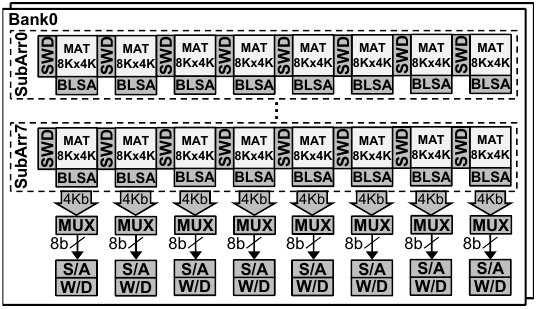}
	\caption{Architecture of a PCM device.}
	\label{fig-media}
\end{figure}

PCM is a non-volatile memory device that has two different states, \textit{amorphous} and \textit{crystalline}. The former has a higher resistance than the latter \cite{RMWMerge}. 
The detailed overview of a PCM device in an 8GB dual-rank module is illustrated in Figure~\ref{fig-media}. The device consists of eight subarrays, and each subarray is composed of eight \rev{cell matrices (MATs)}. Main wordline drivers activate a subarray in each bank. Using the row address, each sub-wordline driver (SWD) activates 4Kb data. The activated data are sensed by bitline sense amplifiers (BLSA) and transferred through global bitlines. Using the column address, each column multiplexer (MUX) outputs an 8-bit word to global sense amplifiers (S/A) by multiplexing 4Kb data. Finally, 8 words are transferred to the data bus in burst mode. In total, 64B are carried out from eight devices, which are driven symmetrically by a single command. For a write operation, data on write drivers (W/D) are written back to the cell array.% with differential write \cite{DCW}. 

\subsection{Modeling Write Disturbance in PCM} \label{S2.2}

WDE is caused by the resistance shift from the amorphous state to the crystalline state \cite{DIN, SDPCM, ModelPCM2}. WDEs occur on an idle cell adjacent to the cell under RESET operations \cite{DIN, SDPCM}. Since the intensity of current during a SET operation is nearly half of that during a RESET operation, an idle cell's temperature next to the programmed cell would be higher than those under SET (but lower than those cells under RESET). As a consequence, a phase transition may occur on that idle cell.

Knowing the occurrence moment of WDEs is also crucial for modeling WDEs in a simulator. Rather than triggering WDEs randomly, the study in \cite{ModelPCM2} explains when a WDE occurs, according to low-level characteristics of WDE. It shows that an amorphous cell gradually shifts to crystalline state due to heat transfer to neighbors, thereby incurring WDEs. The study also explains that a cell can be programmed in different time frames using one pulse per frame; hence, WDE can occur regardless of the idle time duration between consecutive writes. In this paper, we refer the number of pulses incurring WDEs as the \textit{WDE limitation number}.

The prior work in \cite{SIWC} reports the WDE limitation number as 5K-10K, but the author in \cite{SIWC} does not set that number for evaluation. Instead, both this study and \cite{SIWC} assume that a WDE occurs when the number of writes (i.e., 1-to-0 bit flips) exceeds the WDE limitation number of 1K, and this is uniformly applied to all cells. This is because setting the number of 5K or 10K requires a much longer simulation time for triggering WDEs in a row. Our proposed method can be simply extended to various WDE limitation numbers, because the threshold for generating rewrite command is formalized as the function of the WDE limitation number.

Furthermore, the industry has presented that WDEs mainly occur on adjacent materials patterned on a common bitline \cite{DriveReduceWDE}. This is because PCM cells are overlapped with bitlines, incurring simpler heat dissipation along bitlines. Therefore, WDEs mainly occur on adjacent materials patterned on a common bitline. However, our proposed scheme can be easily extended when more than two neighbor cells are disturbed by generating more rewrite operations, which are used for restoring vulnerable cells on demand.

\subsection{Motivation} \label{S2.4}
\textbf{Necessity of reducing the cache burden}. Cache-based schemes mitigate WDEs by temporarily storing write data into dedicated SRAM. Although a cache-based scheme (i.e., \textit{SIWC} \cite{SIWC}) can significantly reduce the number of WDEs in PCM compared to those in previous studies (see Section~\ref{S6.7}), this strategy requires high-capacity SRAM, because it indiscriminately caches write data. Furthermore, data adjacent to cached addresses remain vulnerable to WDEs. To overcome these challenges, it is necessary to store the data that likely incur WDEs (i.e., WDE aggressors) and restore cells adjacent to these aggressors. Therefore, a preprocessor that filters non-aggressors and restores cells adjacent to aggressors is necessary (called “main table”) if a small-sized cache (called “barrier buffer”) is desirable.

\begin{table} [t]
	\footnotesize % \small for ACM
	\caption{Performance of randomized VnC}
	\label{tab-randVnC}
	\centering
	\begin{tabular}{|c|c|c|c|c|}
		\hline
		\multicolumn{3}{|c}{\textbf{Probabilities of VnC}} & \multicolumn{1}{|c}{\textbf{WDE}} & \multicolumn{1}{|c|}{\multirow{2}{*}{\textbf{Speedup}}} \\
		\cline{1-3}
		\multicolumn{1}{|c}{both rows} & \multicolumn{1}{|c}{upper row} & \multicolumn{1}{|c}{lower row} & \multicolumn{1}{|c}{\textbf{reduction}} & \multicolumn{1}{|c|}{} \\
		\hline		
		\multicolumn{1}{|c}{0\%} & \multicolumn{1}{|c}{50\%} & \multicolumn{1}{|c}{50\%} & \multicolumn{1}{|c}{23\%} & \multicolumn{1}{|c|}{57\%} \\
		\hline
		\multicolumn{1}{|c}{75\%} & \multicolumn{1}{|c}{12.5\%} & \multicolumn{1}{|c}{12.5\%} & \multicolumn{1}{|c}{30\%} & \multicolumn{1}{|c|}{18\%} \\
		\hline
		\multicolumn{1}{|c}{80\%} & \multicolumn{1}{|c}{10\%} & \multicolumn{1}{|c}{10\%} & \multicolumn{1}{|c}{36\%} & \multicolumn{1}{|c|}{17\%} \\
		\hline
		\multicolumn{1}{|c}{90\%} & \multicolumn{1}{|c}{5\%} & \multicolumn{1}{|c}{5\%} & \multicolumn{1}{|c}{36\%} & \multicolumn{1}{|c|}{15\%} \\
		\hline
		\multicolumn{1}{|c}{95\%} & \multicolumn{1}{|c}{2.5\%} & \multicolumn{1}{|c}{2.5\%} & \multicolumn{1}{|c}{43\%} & \multicolumn{1}{|c|}{15\%} \\
		\hline
		\multicolumn{1}{|c}{99\%} & \multicolumn{1}{|c}{0.5\%} & \multicolumn{1}{|c}{0.5\%} & \multicolumn{1}{|c}{46\%} & \multicolumn{1}{|c|}{14\%} \\
		\hline
	\end{tabular}
\end{table}

\textbf{Necessity of reducing the performance overhead of VnC}. VnC, the most common solution to WDEs, triggers read operations to read two neighboring data before the objective data is updated. Subsequently, two neighbors are read again after the write operation for verification. Finally, VnC is performed iteratively if WDEs occur on the neighbors, degrading the performance markedly by these read operations. 
A naive approach to reducing the number of such read commands is to perform VnC randomly. Table~\ref{tab-randVnC} shows the WDE reduction rate in comparison with the baseline (i.e., no VnC) and the speedup in comparison with the normal VnC (i.e., always verify both rows). For example, the third row assumes probabilities of this tuple are 80\%, 10\%, and 10\%, respectively.
Random VnC yields a 14\% speedup compared to normal VnC and a WDE reduction rate of 46\% compared to the baseline. This is because PCM does not require a refresh operation by default (or an infrequent refresh compared to DRAM), causing cells scarcely to be restored. In contrast, high speedup (i.e., 57\%) is attainable at the expense of reliability. Moreover, the operations of VnC (i.e., pre-write read, write, and post-write read) are strictly ordered; hence, the speedup is not notable even when a probabilistic approach is applied. Please note that these data are extracted based on the configuration in Section~\ref{S6.1}.
As a result, the VnC-based scheme is unsuitable as a preprocessor (i.e., main table) for the filtering mentioned above. Thus, there is  need for a new on-demand approach that accurately predicts vulnerable patterns and reduces the number of WDEs to a small value comparable to VnC.

\begin{table} [t]
	\footnotesize % \small for ACM
	\caption{Characteristics of representative schemes}
	\label{tab-CmpMethods}
	\centering
	\begin{tabular}{|c|c|c|c|c|}
		\hline
		\multicolumn{1}{|c}{\textbf{Schemes}} &
		\multicolumn{1}{|c}{LAZY \cite{SDPCM}} & \multicolumn{1}{|c}{ADAM \cite{ADAM}} & \multicolumn{1}{|c}{SIWC \cite{SIWC}} & \multicolumn{1}{|c|}{IMDB} \\
		\hline
		\multicolumn{1}{|c}{\textbf{Approach}} & \multicolumn{1}{|c}{VnC} & \multicolumn{1}{|c}{Encode} & \multicolumn{1}{|c}{Cache} & \multicolumn{1}{|c|}{Demand} \\
		\hline
		\multicolumn{1}{|c}{\textbf{WDE}} & \multicolumn{1}{|c}{\multirow{2}{*}{High}} & \multicolumn{1}{|c}{\multirow{2}{*}{Low}} & \multicolumn{1}{|c}{\multirow{2}{*}{Moderate}} & \multicolumn{1}{|c|}{\multirow{2}{*}{High}} \\
		\multicolumn{1}{|c}{\textbf{reduction}} & \multicolumn{1}{|c}{} & \multicolumn{1}{|c}{} & \multicolumn{1}{|c}{} & \multicolumn{1}{|c|}{} \\
		\hline
		\multicolumn{1}{|c}{\textbf{Speed}} & \multicolumn{1}{|c}{Low} & \multicolumn{1}{|c}{Moderate} & \multicolumn{1}{|c}{Moderate} & \multicolumn{1}{|c|}{Moderate} \\
		\hline
		\multicolumn{1}{|c}{\textbf{Energy}} & \multicolumn{1}{|c}{High} & \multicolumn{1}{|c}{Moderate} & \multicolumn{1}{|c}{Moderate} & \multicolumn{1}{|c|}{Moderate} \\
		\hline
		\multicolumn{1}{|c}{\textbf{Storage}} & \multicolumn{1}{|c}{Very large} & \multicolumn{1}{|c}{-} & \multicolumn{1}{|c}{Large} & \multicolumn{1}{|c|}{Small} \\
		\hline
	\end{tabular}
\end{table}

Table~\ref{tab-CmpMethods} shows the relative characteristics of previous WDE mitigation schemes (explained in Section~\ref{S8}) against our proposed method, IMDB. The VnC-based approach (i.e., LAZY or lazy correction) incurs a significant performance and energy overhead due to the increased number of commands for verification. LAZY requires an additional “WDE-free” error correction pointer (ECP) device with a lower density than the normal device \cite{SDPCM}. ADAM only requires simple compression logic without storage resources; however, the mitigation performance is much lower than IMDB due to high dependency on application data patterns. SIWC reduces WDEs moderately by introducing a write cache, which is larger than that of IMDB. Meanwhile, IMDB significantly reduces WDEs to the number similar to the lazy correction by restoring vulnerable data on demand with a small SRAM.

\section{IMDB: In-Module Disturbance Barrier} \label{S3}

\subsection{Architectural Overview} \label{S3.1}
\begin{figure} [t]
\centering
\includegraphics[scale=0.9]{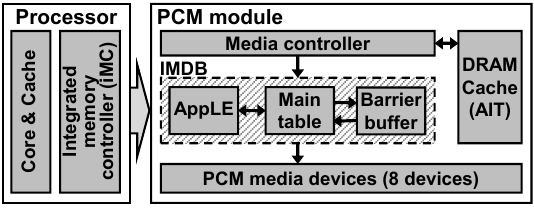}
\caption{Architectural overview of the proposed system. Please note that each PCM media device follows the architecture layout of Figure~\ref{fig-media}.
}
\label{fig-sys}
\end{figure}

Figure~\ref{fig-sys} depicts the overall architecture, where NVM commands are dispatched from the integrated memory controller in the host. For the PCM module, the media controller generates micro-commands and schedules commands to available banks in the media devices. A DRAM cache is only used for storing an address indirection table (AIT) \cite{NVSLOptane, GuidePMEM}. The proposed module, IMDB, is located between the media controller and media devices.

As shown in Figure~\ref{fig-sys}, IMDB consists of the main table (Section~\ref{S3.2.1}), a barrier buffer (Section~\ref{S3.2.2}), and AppLE (Section~\ref{S4.2}). Firstly, the main table manages the addresses of WDE aggressors. If a write address hits in the table, the number of 1-to-0 bit flips is calculated and accumulated in the table; otherwise, the dedicated replacement policy supported by AppLE, which reduces the overhead incurred by multi-port SRAM, selects a victim entry within the table and replaces it with the new address. When the number of bit flips on the aggressor exceeds the pre-defined threshold, IMDB generates \textit{rewrite commands} for data that are adjacent to the aggressor. As explained in Section~\ref{S2.2}, an idle cell in amorphous state (i.e., RESET) gradually shifts to crystalline state if it is exposed to high-temperature several times. Then, a WDE happens when this cell completely turns into crystalline state. Therefore, the rewrite command is introduced and used for restoring such partially shifted cells back to amorphous states before the occurrences of WDEs. Subsequently, IMDB migrates the information from the main table to the barrier buffer that comprises a few data entries, reducing WDEs further. Even though the bit width of a barrier buffer's entry is longer than that of the main table, the barrier buffer manages much fewer entries; hence, it occupies less SRAM capacity than the main table. Figure~\ref{fig-sys} shows the swapping mechanism between two tables, by which WDE aggressors are managed as long as possible within IMDB. 

Our proposed work, IMDB, is a new approach to mitigating WDEs. In particular, IMDB differs from wear leveling and previous WDE studies. 
Wear leveling uniformizes the number of write accesses across different physical regions; however, it just temporally defers WDEs. In contrast, IMDB estimates WDE-vulnerable addresses by utilizing the WDE limitation number and recording aggressors. The estimated vulnerable addresses are then restored to stable states. 
\rev{Furthermore, it is noteworthy that the wear leveling does not affect the threshold selection because the threshold for generating rewrite commands is derived from WDE limitation number, which is determined by the cell characteristics. Indeed, the wear leveling spreads the number of writes over all PCM regions, making it possible to lower occurrences of WDEs \textit{within a fixed time interval}. However, the wear leveling cannot increase the threshold for generating rewrite commands. This is because the wear leveling just temporally postpones occurrences of WDEs. For example, when a data is remapped from cell-A to cell-B due to wear leveling, the state of cell-A remains shifted (i.e., between amorphous and crystalline) because PCM does not require erase operations. Thus, cell-A is still vulnerable to WDEs if another data is mapped to cell-A. It should be noted that WDEs depend on the number of 1-to-0 bit flips on neighboring cells regardless of the rate of programming pulses, as explained in Section~\ref{S2.2}. Consequently, wear leveling is an orthogonal methodology compared to IMDB; wear leveling only defers WDEs rather than reducing WDEs. Furthermore, one of recent studies related to WDEs shows that simply remapping data (e.g., start-gap\cite{SG} or security-refresh\cite{SR}) has small effects on reducing WDEs \cite{WLWD}. On the other hand, IMDB reduces occurrences of WDEs by directly estimating WDE-vulnerable addresses in the main table and barrier buffer.}

In general, there are three categories for mitigating WDEs: VnC-based schemes \cite{VnCinSTT, SDPCM}, encoding schemes \cite{ADAM, WLC, MinWD, ADAPT}, and the cache-based scheme \cite{SIWC}. The VnC-based method defers correction by assuming no error in the additional device; however, VnC basically incurs high performance overhead. In contrast, IMDB restores data before WDEs occur. Encoding schemes are highly dependent on application data patterns. Compared with this kind of schemes, IMDB monitors the vulnerability of data patterns,  leading to less dependence on application data pattern. The cache-based scheme requires a larger SRAM for notably reducing WDEs. On the other hand, IMDB buffers urgent data using the WDE limitation number, reducing WDEs significantly with an SRAM capacity that is four times smaller than the previous study.

\subsection{Implementation of Data Structures} \label{S3.2}

\begin{figure}[t]
\centering
\includegraphics[scale=0.65]{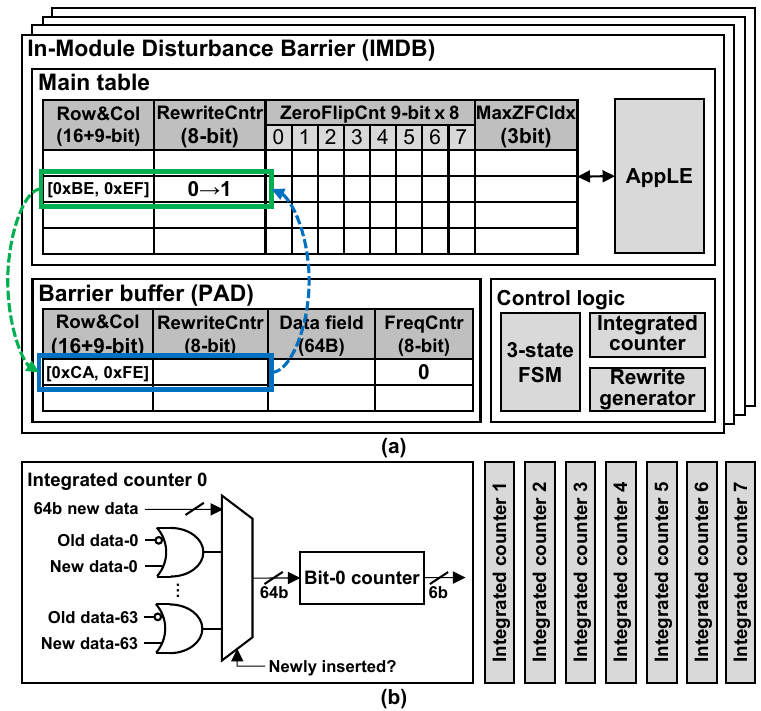}
\caption{Detailed design of IMDB: (a) implementation of four IMDB planes (each IMDB plane is assigned to each PCM bank operation), (b) integrated counters for eight \textit{ZeroFlipCntr}s.}
\label{fig-IMDB}
\end{figure}

Figure~\ref{fig-IMDB}(a) shows the detailed architecture of IMDB, where each plane is allocated for every PCM bank; hence, all IMDB planes operate concurrently at the bank level without contention. An IMDB plane consists of two tables, namely a \textit{main table} and a \textit{barrier buffer}. The following subsections describe implementations of each table.

\subsubsection{Main Table} \label{S3.2.1}
The main table is implemented with a set of SRAMs, where the entry is updated by a control logic. In particular, four fields in the table are used for estimating the degree of WDE on write addresses:

\begin{itemize}
\item \textit{Row \& Col}: Indicates row and column addresses that are currently being managed.
\item \textit{ZeroFlipCntr}: Eight sub-counters are in the field, each of which counts the number of bit flips from 1 to 0 and manages one 64-bit word in a 64B cache line. Each of the eight ZeroFlipCntrs manages a 64-bit word within a device (or chip), because one 64-bit word outputs from each of eight devices, as shown in Figure~\ref{fig-media}. Consequently, these eight ZeroFlipCntrs map to a row of a subarray and symmetrically manage eight 8-bit sub-words across eight devices. 
\item \textit{MaxZFCIdx}: Indicates the sub-counter index of \textit{ZeroFlipCntr} holding the maximum value. It is updated in control logic after reading an entry. It is used for comparing the maximum value of the \textit{ZeroFlipCntr} with the threshold value for rewrite operations.
\item \textit{RewriteCntr}: Represents the frequency of rewrite operations on the address of \textit{Row \& Col} in an 8-bit counter.
\end{itemize}

A per-bank IMDB plane is assigned to each bank; hence, bank parallelism is ensured to lower the contention on IMDB. Furthermore, IMDB prevents resource redundancy, because only one command processed in IMDB at a time without incorporating a serialized queue. The command is handled by a three-state finite state machine (i.e., IDLE, HIT, MISS) in control logic, where the varying latency of the multiple states are factored in the simulator. After a command is inserted, IMDB operates in two different ways, depending on the existence of the address in the table:

\begin{itemize}
\item If the address is found in the main table, the state transits to HIT. Meanwhile, two types of data, i.e., the new write-data and the previously written data already read in the controller, are passed to control logic. Subsequently, the number of 1-to-0 bit flips is counted by integrated counters (see Section~\ref{S4.2}) and accumulated to the corresponding \textit{ZeroFlipCntr}. When the maximum value of \textit{ZeroFlipCntr} surpasses the predefined \textit{threshold}, two rewrites on adjacent wordlines are generated and sent to the write queue in the media controller. Accordingly, the value of \textit{RewriteCntr} increases. 
\item If an address is not found in the main table, an insertion is required while converting the state to MISS. The probabilistic insertion method is leveraged in this study, where infrequent accesses are filtered out with probability \textit{p} to reduce evictions from the SRAM. When insertion is required, our proposed replacement policy determines the victim (explained in Section~\ref{S4}), and thereby the new address can replace the victim entry.
\end{itemize}

According to hit/miss cases on the main table, the finite state machine is a trigger for different operations. For both cases, after the table reference, write data issued to the media right away. Memory commands in the media controller scheduler must follow the promised timing constraints. Thus, no command can be entered to the same IMDB plane during the write phase in the media, allowing the background processing of IMDB.

In the proposed design, two parameters, (1) the threshold of generating rewrite commands and (2) the probability \textit{p}, are necessary. First of all, we decide the threshold of generating rewrite commands in the main table as \textit{WDE limitation number/2-1}, because two rows can disturb a row. Thus, if we assume a WDE limitation number of 1K, as in\cite{SIWC}, the threshold becomes 511, making the bit width of each \textit{ZeroFlipCntr} to be 9. 
The other parameter, \textit{p}, indicates the probability of inserting a new missed address into the main table. Increasing the probability incurs more frequent entry replacement in the table for detecting WDE aggressor, losing the opportunity to rewrite the victims of WDEs. In contrast, lowering the probability makes “long-term” attacks lose the chance to be in the table. Our experiments regarding different insertion probabilities show that \textit{p}=1/128 yields the fewest WDEs; hence, we select \textit{p}=\textbf{1/128}.

As shown in Figure~\ref{fig-IMDB}(a), the main table employs two types of SRAMs. First, a dual-port content-addressable SRAM (CAM) is allocated as \textit{Row \& Col} fields. Second, a multi-port SRAM, consisting of \textit{ZeroFlipCntr}, \textit{MaxZFCIdx}, and \textit{RewriteCntr}, has multiple read ports for obtaining all entry contents at once to apply the proposed replacement policy (see Section~\ref{S4.1}). However, since the use of multi-port SRAMs causes a significant overhead, we propose \textit{AppLE}, which enables the replacement policy with a DPSRAM without speed degradation (see Section~\ref{S4.2}).

\begin{figure} [t]
\centering
\includegraphics[scale=0.9]{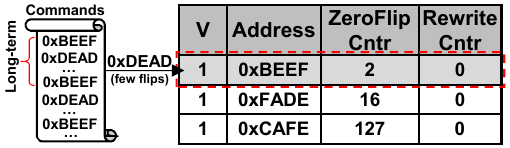}
\caption{A toy example showing malicious attacks. 0xDEAD evicts insufficiently baked 0xBEEF, which is vulnerable to WDEs with gradual 1-to-0 bit flips.}
\label{fig-malicious}
\end{figure}

\subsubsection{Barrier Buffer} \label{S3.2.2}
The barrier buffer is introduced to store the data with frequent 1-to-0 bit flips. For a read request, the barrier buffer is capable of serving commands directly. For a write command, if the address hits on the barrier buffer, the data are updated in the barrier buffer directly. Otherwise, if an address hits only on the main table, the normal operation of the main table is performed, as explained in Section~\ref{S3.2.1}.

As shown in Figure~\ref{fig-IMDB}(a), the green-boxed entry in the main table contains the data frequently exposed to 1-to-0 flips. It is invalidated and \textit{promoted} to the barrier buffer when \textit{RewriteCntr} updates (i.e., rewrite occurs in the main table). The barrier buffer inherits the address and \textit{RewriteCntr} information from the main table. If the barrier buffer is not full, the promoted entry can be directly placed in the barrier buffer. 
\rev{After several entry promotions (i.e., rewrite operations) from the main table, the barrier buffer would become full. At this moment, the promoted entry (from the main table) replaces the least frequently used (LFU) entry that is bounded by the blue box in Figure~\ref{fig-IMDB} (a). For this reason, \textit{FreqCntr} is required for the replacement policy, as in \cite{PAD}. The LFU entry data are then sent back to the media controller for writing back the dirty data, and this information is demoted to the main table. }
Because the demoted addresses have been WDE aggressors before, the number of rewrites is reserved in \textit{RewriteCntr}. \textit{RewriteCntr} provides historical information with which to obtain a reasonable victim candidate in the main table (explained in Section~\ref{S4.1}). Please note that the 8-bit of \textit{RewriteCntr} is a generously selected bit width to prevent overflow based on our experiments.

To implement the barrier buffer, a dual-port CAM-based SRAM and a dual-port SRAM are employed for \textit{Row \& Col} and \textit{data \& RewriteCntr \& \it FreqCntr}, respectively. The energy consumption is negligible, because only a small number of entries in the barrier buffer are necessary to provide high WDE mitigation performance, as shown in Section~\ref{S6.7.3}. The sensitivity analysis of the number of entries will be shown in Section~\ref{S6.5}.

\subsection{Modification of Media Controller} \label{S3.3}
\rev{The media controller is modified to support IMDB in two aspects. First, acquiring the old data is necessary to count bit flips. Thus, a \textit{pre-write read operation} is performed ahead of a write command. The pre-write read request has a higher priority than write requests but a lower priority than normal read requests because write requests in the controller mainly drain when the queue is full. Lastly, a merge operation is introduced, by which the rewrite command can coalesce with a same-address write command. }

\begin{figure} [t]
\centering
\includegraphics[scale=0.75]{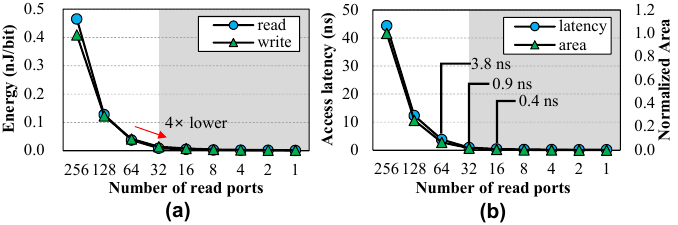}
\caption{Characteristics of a 256-entry SRAM having multiple read ports, which is extracted from CACTI \cite{CACTI}: (a) energy, (b) latency and area.}
\label{fig-SRAMports}
\end{figure}

\section{Replacement Policy} \label{S4}

\subsection{Replacement Policy for IMDB} \label{S4.1}
A replacement (or eviction) policy is required in the main table based on the characteristics of WDEs. Therefore, we exploit \textit{ZeroFlipCntr} and \textit{RewriteCntr} to define the replacement policy. 
When the input command requests a new entry in the main table, the policy is ready to select the victim entry. The victim candidate is defined as a less urgent aggressor, thereby selecting the minimum value of \textit{ZeroFlipCntr}. However, more than two candidates may exist if the table has multiple entries with the same values of \textit{ZeroFlipCntr}. Since the aggressiveness of WDEs varies with historical information (i.e., \textit{RewriteCntr}), the entry containing the minimum of \textit{RewriteCntr} is finally selected as the replaced entry.

To prevent "cold-start" that incurs early eviction from the table, this study introduces \textit{prior knowledge}. Since the policy prioritizes the present vulnerability using \textit{ZeroFlipCntr}, the recently inserted but insufficiently "baked" entry can easily be evicted from the main table. Although \textit{RewriteCntr} contains the historical information, it becomes useless if the entry is newly inserted and evicted right away (see example in Figure~\ref{fig-malicious}). To tackle this problem, the prior knowledge, which is simply defined as the number of zeros in each data block, is stored in \textit{ZeroFlipCntr}. 

It is noteworthy that a module, namely \textit{integrated counter}, is required to perform the above processes. The integrated counter provides mainly two functions. First, it counts the number of 0s of newly inserted data, which is then directly used as \textit{prior knowledge} of \textit{ZeroFlipCntr}. Second, it counts the number of 1-to-0 bit flips of the accessed address in the table. The counted value is then added to the \textit{ZeroFlipCntr}. As a result, the integrated counter is implemented as Figure~\ref{fig-IMDB}(b), where eight counter blocks are required to count each 64-bit word in a 64-byte data concurrently. 

\subsection{Approximate Lowest Number Estimator} \label{S4.2}
The eviction policy requires the number of read ports to be equal to the number of entries on the main table. It increases latency, area, and energy overheads. If a 256-entry main table is assumed, 255 tree-structured dual-input comparators are necessary for latency minimization (i.e., 8 cycles). However, our evaluation results in Figures~\ref{fig-SRAMports} (a) and (b) indicate that increasing the number of read ports on an SRAM significantly increase overheads. As a result, an SRAM with 256 read ports is an infeasible implementation. 

To reduce such overheads, this paper introduces a sampling-based comparator, called AppLE. The basic concept of AppLE is to bind multiple entries. For example, binding 8 entries results in 32 groups.
In this case, a randomly generated number ranging from 0 to 7 is multiplied by 8 and assigned to each group (i.e., \textit{group-index$\times$8}). This assigned value is used as the main table's input address, and a \textit{sampled entry} is referenced. Using this addressing mechanism, the victim candidate is selected among sampled entries.

\begin{figure} [t]
\centering
\includegraphics[scale=0.7]{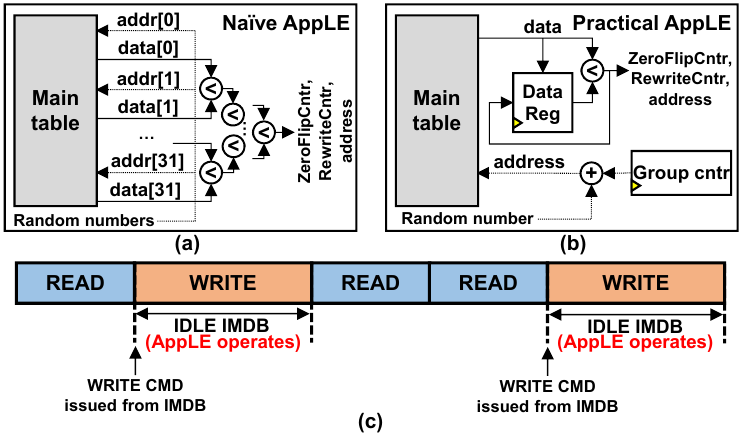}
\caption{Implementations of AppLE: (a) a naive approach, (b) a practical approach, (c) timeline of IDLE state in IMDB.}
\label{fig-AppLE}
\end{figure}

The main concept of AppLE is comparing counts approximately by grouping a few entries in the main table, instead of comparing counts in parallel. Two design options are first discussed in this paper: the first one naively implements approximate counting in parallel (Figure~\ref{fig-AppLE} (a)); the second one performs approximate counting sequentially without increasing the latency on the critical path (Figure~\ref{fig-AppLE} (b)). The first option (i.e., the parallel one) is infeasible to be implemented in the industry, because it simply regards the number of groups as the number of read ports. For example, the typical I/O frequency of DDR4 is around 800MHz \cite{DDR4wiki}, and the maximum target number of read ports is set to 32. Still, the area of a 32-port SRAM is 105$\times$ larger than that of a single read port SRAM. Moreover, an SRAM consisting of dozens of read ports is unusual in terms of manufacturing. 

This is the reason for choosing the second design option. In the second design, the latency of sequential comparisons can be hidden within the IDLE state. This is because the number of comparisons reduces with AppLE (e.g., 32 cycles for above example), and the IDLE state maintains for 120 cycles after issuing a write command. 
We directly evaluate the case of comparing all 256 entries (i.e., no-AppLE) in Figure~\ref{fig-ECC} (b); it shows no-AppLE case incurs 15\% of performance degradation, because 136 cycles (=256-120) of additional latency cannot be hidden within the IDLE state. 

\begin{table} [t]
	\footnotesize % \small for ACM
	\caption{Simulation configurations}
	\label{tab-cfg}
	\centering
	\begin{tabular}{|c|c|c|}
		\hline
		\multicolumn{1}{|c}{\textbf{Simulator}} & \multicolumn{1}{|c}{\textbf{Device}} & \multicolumn{1}{|c|}{\textbf{Description}} \\
		\hline
		\hline 
		\multicolumn{1}{|c}{\multirow{6}{*}{gem5}} & \multicolumn{1}{|c}{Cores} & \multicolumn{1}{|l|}{Out-of-order, 4-core, 2GHz} \\
		\cline{2-3}
		\multicolumn{1}{|c}{} & \multicolumn{1}{|c}{\multirow{3}{*}{L1 cache}} & \multicolumn{1}{|l|}{I-cache: 2-way set associative,} \\
		\multicolumn{1}{|c}{} & \multicolumn{1}{|c}{} & \multicolumn{1}{|l|}{D-cache: 4-way set associative,} \\
		\multicolumn{1}{|c}{} & \multicolumn{1}{|c}{} & \multicolumn{1}{|l|}{each has a capacity of 64KB.} \\
		\cline{2-3}
		\multicolumn{1}{|c}{} & \multicolumn{1}{|c}{\multirow{2}{*}{L2 cache}} & \multicolumn{1}{|l|}{Shared last-level cache. 16-way} \\
		\multicolumn{1}{|c}{\multirow{7}{*}{NVMain}} & \multicolumn{1}{|c}{} & \multicolumn{1}{|l|}{set associative, 1MB.} \\
		\hline
		\multicolumn{1}{|c}{} & \multicolumn{1}{|c}{Media} & \multicolumn{1}{|l|}{Separated write queue and read} \\
		\multicolumn{1}{|c}{} & \multicolumn{1}{|c}{controller} & \multicolumn{1}{|l|}{queue (64-entry), FR-FCFS.} \\
		\cline{2-3}
%		\multicolumn{1}{|c}{} & \multicolumn{1}{|c}{\multirow{4}{*}{PCM}} & \multicolumn{1}{|l|}{Differential write supported} \\
		\multicolumn{1}{|c}{} & \multicolumn{1}{|c}{\multirow{3}{*}{PCM}} & \multicolumn{1}{|l|}{Read: 100ns, RESET: 100ns, SET: 150ns} \\
		%\multicolumn{1}{|c}{} & \multicolumn{1}{|c}{} & \multicolumn{1}{|l|}{RESET: 100ns, SET: 150ns} \\
		\multicolumn{1}{|c}{} & \multicolumn{1}{|c}{} & \multicolumn{1}{|l|}{Write disturbance limitation: 1K} \\
		\multicolumn{1}{|c}{} & \multicolumn{1}{|c}{} & \multicolumn{1}{|l|}{Size: 8GB (2-rank, 2-bank/rank)} \\
		\hline
	\end{tabular}
\end{table}

\section{Evaluations} \label{S6}

\subsection{Configurations} \label{S6.1}

Table~\ref{tab-cfg} shows the configuration of evaluation environment. 
\rev{In this study, we use four simulators to simulate a PCM-based main memory system: gem5\cite{GEM5}, NVMain\cite{NVMain2}, NVSim\cite{NVSim}, and CACTI\cite{CACTI}. Gem5 is a processor architecture simulator that is configured as a quad-core processor \cite{GEM5}. NVMain is a simulator that simulates details of NVM subsystems \cite{NVMain2}. Both simulators are functional- and cycle-accurate; hence, running gem5 and NVMain together requires extremely long simulation time. Moreover, sensitivity analysis requires more than 400 experiments in this paper. Thus, trace-driven simulation is necessary to significantly reduce the simulation time. Trace-driven simulation is a common evaluation methodology in NVM-related studies, as performed in \cite{MinWD} and \cite{ADAM}. To conduct the trace-driven simulation, we first extract memory command traces by running workloads on gem5 in standalone mode. Thereafter, extracted command traces are fed into NVMain, which can also be run in a standalone manner. NVSim \cite{NVSim} and CACTI \cite{CACTI} are energy simulators to estimate energy parameters (i.e., energy per access) of PCM and SRAM. The energy evaluation mechanism in NVMain calculates the energy consumption of two memory types using energy parameters obtained from these two energy simulators.}
Still, a large L2 cache in the processor requires a long simulation time to incur enough WDEs (i.e., more than 100); hence, it is necessary to determine a small but practical L2 cache size to build a burn-in test environment. Therefore, the processor is configured as the mobile processor \cite{MobileCPU}, which may incur increased memory traffic. Nonetheless, it should be noted that we extract memory traces having a wide range of misses per thousand instructions (MPKI) in order to simulate the various kinds of memory traffic, as shown in Table~\ref{tab-WL}. In this study, traces are obtained from SPEC CPU benchmark suit \cite{SPECPU2006} and synthesized persistent workloads (prefixed as “pmix”) that are similar to those in \cite{NVHeaps, ATOM, REDU}. Please note that the \textbf{baseline} does not apply any mitigation scheme.

\begin{table} [t]
	\footnotesize % \small for ACM
	\caption{Information on workloads}
	\label{tab-WL}
	\centering
	\begin{tabular}{|c|c|c|}
		\hline
		\multicolumn{1}{|c}{\textbf{Workloads}} & \multicolumn{1}{|c}{\textbf{Description}} & \multicolumn{1}{|c|}{\textbf{MPKI}} \\
		\hline
		\hline
		\multicolumn{1}{|c}{SPEC::bzip2} & \multicolumn{1}{|l}{General compression} & \multicolumn{1}{|c|}{11.98} \\
		\hline
		\multicolumn{1}{|c}{SPEC::sjeng} & \multicolumn{1}{|l}{Artificial intelligence (chess)} & \multicolumn{1}{|c|}{0.89} \\
		\hline
		\multicolumn{1}{|c}{SPEC::h264ref} & \multicolumn{1}{|l}{Video compression} & \multicolumn{1}{|c|}{1.65} \\
		\hline
		\multicolumn{1}{|c}{SPEC::gromacs} & \multicolumn{1}{|l}{Biochemistry} & \multicolumn{1}{|c|}{5.49} \\
		\hline
		\multicolumn{1}{|c}{SPEC::gobmk} & \multicolumn{1}{|l}{Artificial intelligence (go)} & \multicolumn{1}{|c|}{6.65} \\
		\hline
		\multicolumn{1}{|c}{SPEC::namd} & \multicolumn{1}{|l}{Biology} & \multicolumn{1}{|c|}{1.09} \\
		\hline
		\multicolumn{1}{|c}{SPEC::omnetpp} & \multicolumn{1}{|l}{Discrete event simulation program} & \multicolumn{1}{|c|}{6.99} \\
		\hline
		\multicolumn{1}{|c}{SPEC::soplex} & \multicolumn{1}{|l}{Linear programming optimization} & \multicolumn{1}{|c|}{21.31} \\
		\hline
		\multicolumn{1}{|c}{pmix1} & \multicolumn{1}{|l}{Queue, Hashmap, B-tree, Skiplist} & \multicolumn{1}{|c|}{10.24}\\
		\hline		
		\multicolumn{1}{|c}{pmix2} & \multicolumn{1}{|l}{Queue, B-tree, RB-tree, Skiplist} & \multicolumn{1}{|c|}{11.10}\\
		\hline		
		\multicolumn{1}{|c}{pmix3} & \multicolumn{1}{|l}{Hashmap, RB-tree, Queue, Skiplist} & \multicolumn{1}{|c|}{8.95}\\
		\hline
		\multicolumn{1}{|c}{pmix4} & \multicolumn{1}{|l}{RB-tree, Hashmap, B-tree, Skiplist} & \multicolumn{1}{|c|}{10.12}\\
		\hline
	\end{tabular}
\end{table}

\subsection{Architectural Exploration} \label{S6.2}
Design parameters, specifically the number of entries in the main table ($N_{mt}$), the number of entries in the barrier buffer ($N_{b}$), and the group size dedicated to AppLE ($N_{g}$), are crucial when seeking a cost-effective architecture for IMDB. As explained in the previous section, the latency of AppLE can be entirely hidden by the IDLE state of IMDB from $N_{g}=$32 (see Figure~\ref{fig-SRAMports}), which also holds for $N_{g}<$32. Moreover, 64 is determined as the maximum number of entries in the barrier buffer to guarantee that no more than 10\% of the flush time (i.e., 100us) is consumed. As a result, the trade-off function of IMDB is defined as follows: 
%\newline

\begin{equation} \label{eq-cost}
\begin{split}
T=W(N_{mt}, N_{b}, N_{g})+A(N_{mt}, N_{b})+S^{-1}(N_{b}), \\ 
where~N_{g}\leq32, N_{b}\leq64
\end{split}
\end{equation}
\newline 
where $W$, $A$, and $S$ are the number of WDEs, the area, and the speedup (i.e., execution time normalized to the baseline \cite{SDPCM}), respectively. Based on Eq~(\ref{eq-cost}), this section evaluates the effectiveness of the prior knowledge and determines the main table size ($N_{mt}$). Subsequently, sensitivity analyses concerning the number of entries in the barrier buffer ($N_{b}$) and the group size for AppLE ($N_{g}$) are conducted to determine the cost-effective parameters. Finally, these parameters are applied and compared to previous studies.

\subsection{Effectiveness of the Replacement Policy} \label{S6.3}

\begin{figure} [t]
\centering
\includegraphics[scale=0.75]{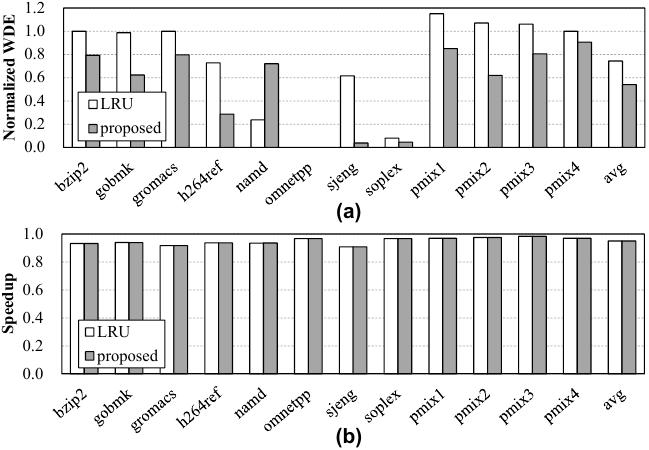}
\caption{Performance according to different replacement polices: (a) normalized WDE, (b) speedup.}
\label{fig-policy}
\end{figure}

Several replacement policies have been published in previous studies, such as MRU (most-recently used), LFU (least frequently used), and LRU-alike policies (e.g., pseudo-LRU). It is noteworthy that WDEs occur when a neighboring cell is frequently programmed. Thus, we need to consider this characteristic when choosing the appropriate policy.  MRU discards the most recently used items. However, WDEs may occur on some applications with relatively high locality. For LFU, we need to add additional metadata on the entry for representing the access frequency, incurring higher resource costs. As a result, we finally compare the proposed policy against the LRU, because LRU simultaneously considers the locality and the access frequency.
Figure~\ref{fig-policy}(a) shows that the LRU yields higher WDEs than the proposed policy, because the LRU makes the address close to WDEs be evicted if it is not accessed for a long time. For example, \textit{bzip2, gobmk, gromacs}, and persistent workloads have this kind of access pattern, increasing WDEs. In contrast, the proposed policy observes the number of bit flips and keeps track of their long-term history. However, the LRU shows 3$\times$ fewer WDEs than the proposed policy on \textit{namd}. This is because \textit{namd} has high spatial and temporal locality. We find that \textit{namd} achieves a 70\% higher row buffer hit rate than an application of a similar MPKI (i.e., \textit{sjeng}), yielding lower hit rate on the main table. However, such a degradation will be mitigated in the following subsections. 

Figure~\ref{fig-policy} (b) shows that replacement policies for the main table do not affect the speed performance, because the main functionality of the IMDB is managing WDE aggressors without caching plenty of data in SRAM. On the other hand, the proposed policy generally contributes to lower WDEs (see Figure~\ref{fig-policy} (a)), because it keeps aggressors more precisely than the LRU and rewrites rows adjacent to aggressors.

\subsection{Sensitivity to Main Table Configuration} \label{S6.4}

\begin{figure} [t]
\centering
\includegraphics[scale=0.75]{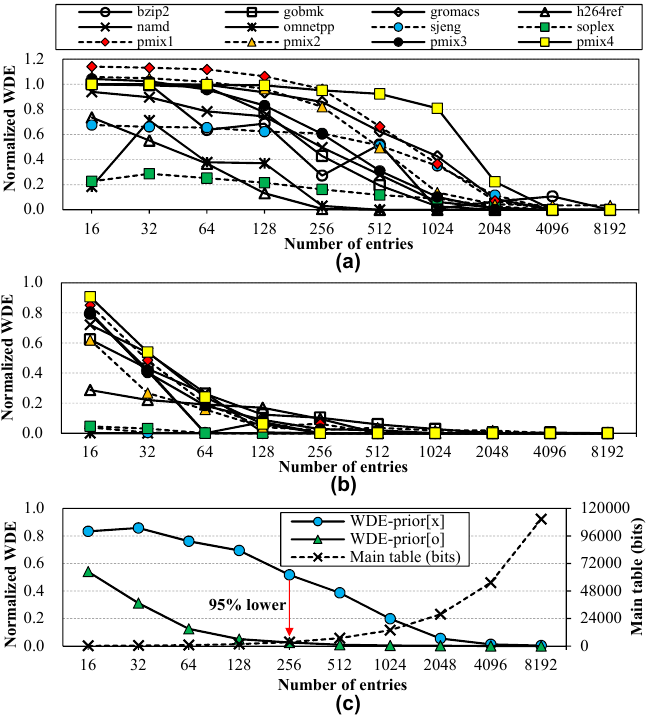}
\caption{Normalized WDEs regarding different numbers of entries in the main table, (a) without prior knowledge, (b) with prior knowledge, (c) displaying average normalized WDE and SRAM capacity.}
\label{fig-prior-WDE}
\end{figure}

Figures~\ref{fig-prior-WDE}(a) and (b) show the normalized WDE regarding different numbers of entries in the main table. Both figures show that WDEs generally decrease as the number of entries increases. In particular, as shown in Figure~\ref{fig-prior-WDE}(a), while the number of WDEs exceeds that in the baseline when the number of entries is fewer than 256, the number decreases sharply from 2048 entries.
This is because the small-size table cannot be trained due to frequent entry replacement on the main table.
On the other hand, as shown in Figure~\ref{fig-prior-WDE}(b), the 256-entry main table with prior knowledge yields a result equivalent to that of the 2048-entry table without prior knowledge. In other words, the proposed method yields an eightfold increase in the efficiency of the WDE mitigation performance.

Figure~\ref{fig-prior-WDE}(c) presents the average normalized WDE and the capacity required for the main table, and the probabilistic insertion scheme discussed in Section~\ref{S3.2.1} is already adopted for both configurations. As shown in Figure~\ref{fig-prior-WDE}(c), the normalized WDE is \textbf{95\%} lower than the case without prior knowledge at 256 entries. Furthermore, the main table's capacity significantly increases from 512 entries; hence, 256 entries can be selected as an appropriate number of entries in the main table, considering the trade-off between the performance and the area. In summary, from this subsection, the number of entries in the main table is fixed as \textbf{$N_{mt}=$256}.

\begin{figure} [t]
\centering
\includegraphics[scale=0.75]{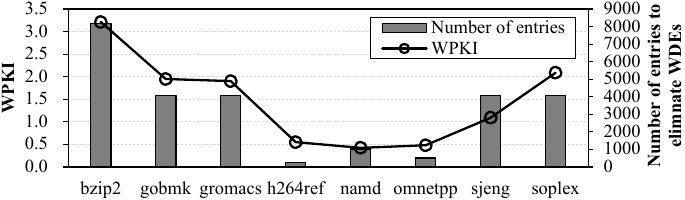}
\caption{Relationship between WPKI and the number of main table entries.}
\label{fig-WPKI}
\end{figure}

\rev{Rather than write request rates or write patterns (e.g., stride or stream), the number of WDEs fundamentally relies on the number of 1-to-0 bit flips on neighboring addresses. In other words, data programming patterns of applications determine the overall WDE occurrences in a PCM device. However, we find that the number of writes per thousand instructions (WPKI) determines the number of main table entries. Figure~\ref{fig-WPKI} shows the relationship between WPKI and the number of main table entries; the number of table entries in this figure denotes the number of entries to fully eliminate WDEs only using the main table. In general, fewer entries are required for smaller WPKI values because WPKI determines the footprint of write commands. Moreover, a larger footprint of write commands incurs more frequent replacements on the main table, yielding more WDEs. Therefore, the main table needs to operate with the barrier buffer and AppLE for higher mitigation performance, because each component of IMDB is complementary to each other. 
}

\subsection{Sensitivity to Barrier Buffer Size} \label{S6.5}

\begin{figure} [t]
\centering
\includegraphics[scale=0.75]{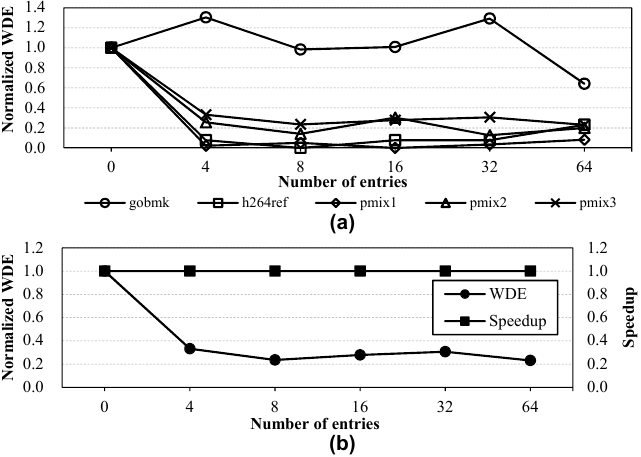}
\caption{Sensitivity to the number of entries in barrier buffer: (a) normalized WDE, (b) average performance.}
\label{fig-barrier-perf}
\end{figure}

Figure~\ref{fig-barrier-perf}(a) shows the number of WDEs with different numbers of entries (i.e., different sizes) in the barrier buffer. For clarity, the results are normalized to the \textit{temporal base condition}; that is, the main table consists of 256 entries with the prior knowledge. Please note that Figure~\ref{fig-barrier-perf}(a) only shows benchmarks still having WDEs under the temporal base condition. As shown in this figure, most benchmarks yield significantly fewer WDEs with the 4-entry barrier buffer. On the other hand, WDEs in \textit{gobmk} decrease when the 64-entry is applied, because some write patterns have extremely long period; these are unlikely to be affected by the proposed policy regardless of the buffer size. However, the following subsection shows that AppLE resolves this problem.

Figure~\ref{fig-barrier-perf}(b) shows the average normalized WDE of the benchmarks mentioned above. Because the speedup does not increase remarkably considering the number of entries, $S^{-1}$ in Eq~(\ref{eq-cost}) can be referred to as a constant. Furthermore, the capacity of the barrier buffer is at least three times as small as the main table for $N_{b}\leq$16 (see bit widths of tables in Figure~\ref{fig-IMDB}(a)), which makes the capacity of the barrier buffer negligible compared to the main table. 
It can be concluded that $W$ in Eq~(\ref{eq-cost}) is sufficient to obtain a cost-effective architecture. Therefore, we select \textbf{$N_{b}=$8} as the trade-off point, because the WDE stabilizes from 8 entries (i.e., \textbf{76.5\%}).

\subsection{Sensitivity to AppLE Group Size} \label{S6.6}

\begin{figure} [t]
\centering
\includegraphics[scale=0.75]{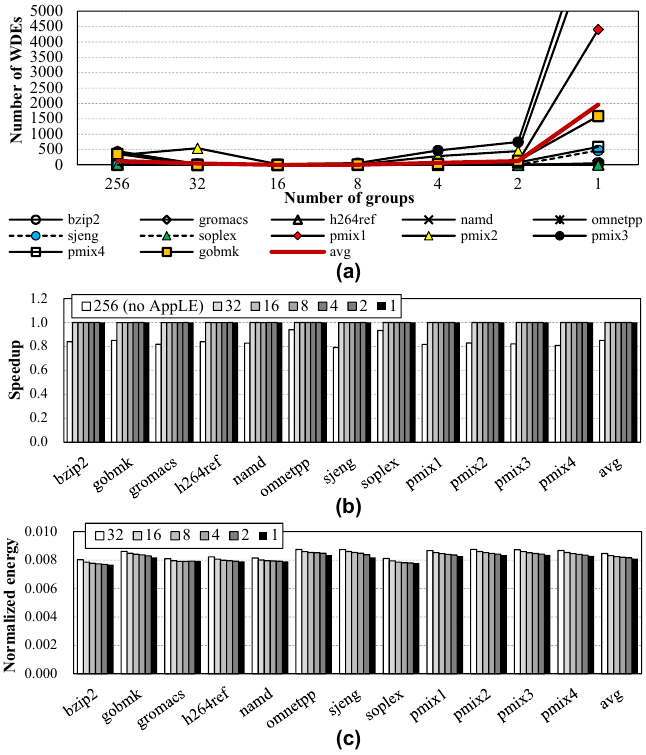}
\caption{Sensitivity to the number of groups in AppLE: (a) normalized WDE, (b) speedup, (c) normalized energy.}
\label{fig-AppLE-perf}
\end{figure}

Figure~\ref{fig-AppLE-perf}(a) presents the absolute number of WDEs with different numbers of groups. Here, the barrier buffer is not applied for straightforward analysis, and 256 groups mean that AppLE is not applied. As presented in Figure~\ref{fig-AppLE-perf}(a), WDEs lower with fewer groups for most benchmarks. Furthermore, AppLE has the potential for avoiding “tricky patterns”. The worst-case behavior for WDEs can be caused by repetitive 0 and 1 pulses on the same address, which incurs WDEs on 512$\times$2=1024 bits. However, the main table can easily detect such a pattern, because it manages the number of 1-to-0 bit flips and generates rewrite operations on vulnerable addresses. In contrast, a trickier way to induce WDEs is incurring 1-to-0 bit flips on an address (say "A") with a long period (e.g., \textit{gobmk}). Furthermore, a large number of unrepeated addresses except "A" are programmed in this long period (i.e., ABC...ADE...A...). This tricky pattern confuses the main table and frequently replaces entries; however, AppLE binds multiple entries as a group, and only one entry randomly becomes a replacement candidate within a group. Therefore, the adversarial address rarely gets evicted from the table for a larger group size. The graph of \textit{gobmk} in Figure~\ref{fig-AppLE-perf} (a) shows that the group size of 8 (whereby the number of groups is 32) yields lower WDEs than the case without AppLE. However, WDEs increase significantly from 2 groups (see red graph in Figure~\ref{fig-AppLE-perf} (a)). In particular, the \textit{fully randomized replacement policy} (i.e., one group) shows 15$\times$ more WDEs than the case without AppLE, indicating that the fully randomized replacement policy is less reliable. As a result, \textbf{$N_{g}=$8} or \textbf{4} is selected as an appropriate design parameter for AppLE.

Figure~\ref{fig-AppLE-perf}(b) presents speedups regarding different numbers of groups. If AppLE is not applied, a 256-cycle latency is induced at least. Even though the latency can be hidden within the write latency (i.e., 120 cycles), at least 136 remaining cycles slow down the performance by 15\%, as shown in the figure. In contrast, AppLE has no performance degradation due to latency hiding. Regarding energy consumption, Figure~\ref{fig-AppLE-perf}(c) shows the SRAM energy normalized to the case without AppLE. In general, the energy decreases as the number of read ports shrinks. 

\subsection{Comparison with Other Studies} \label{S6.7}
% WDE, Speedup, Energy

From the sensitivity analysis above, the most cost-effective IMDB becomes IMDB(e256b8g8), which consists of 256 entries in the main table, 8 entries in the barrier buffer, and a group size of 8. The group size of 4 is denoted as IMDB (e256b8g4). Five schemes are compared against IMDB: (1) \textit{PARR}, (2) \textit{FnW} \cite{FlipNWrite}, (3) \textit{Lazy correction} \cite{SDPCM}, (4) \textit{ADAM} \cite{ADAM}, and (5) \textit{SIWC} \cite{SIWC}. 

PARA (probabilistic adjacent row activation) is commonly used for mitigating rowhammers in DRAM devices \cite{DRAMDisturb}. Preventing occurrences of WDEs requires restoration (i.e., rewrite) rather than activation; hence, rewrite commands for adjacent row data should be randomly generated when a normal write command goes into the media controller. This study evaluates \textit{PARR}s (probabilistic adjacent row restoration) with different probabilities (i.e., p=0.1-0.0001). \textit{FnW} inverts the data if more than half of the bits are changed \cite{FlipNWrite}; it can minimize the number of bit flips in a PCM device. Since \textit{FnW} is a device-level approach, it is applied to the proposed scheme. \textit{Lazy correction} defers subsequent VnC by temporarily storing errors in an ECP chip \cite{SDPCM}. Each entry of ECP records multiple errors of one PCM line. We assume that 10 pointers, which is the maximum number in \cite{SDPCM}, are handled in the ECP. \textit{ADAM} aligns the compressed data in the device alternately to avoid data pattern that is vulnerable to WDEs \cite{ADAM}. \textit{SIWC} sparsely caches write data in an SRAM \cite{SIWC}. In particular, \textit{SIWC-size} indicates that the SRAM capacity is identical to that of IMDB, and \textit{SIWC-entry} holds entries in an amount equal to that of IMDB.

\begin{table} [t]
	\small
	\caption{Performance of different mitigation schemes}
	\label{tab-avg}
	\centering
	\begin{tabular}{|c|c|c|c|}
		\hline
		\multicolumn{1}{|c}{\textbf{Schemes}} & \multicolumn{1}{|c}{\textbf{WDEs}} & \multicolumn{1}{|c}{\textbf{Speedup}} & \multicolumn{1}{|c|}{\textbf{Energy}} \\
		\hline
		\hline
		\multicolumn{1}{|c}{PARR(p=0.1)} & \multicolumn{1}{|c}{41.915} & \multicolumn{1}{|c}{0.9718} & \multicolumn{1}{|c|}{1.09468} \\
		\hline
		\multicolumn{1}{|c}{PARR(p=0.01)} & \multicolumn{1}{|c}{4.5090} & \multicolumn{1}{|c}{0.9971} & \multicolumn{1}{|c|}{1.00947} \\
		\hline
		\multicolumn{1}{|c}{PARR(p=0.001)} & \multicolumn{1}{|c}{0.2670} & \multicolumn{1}{|c}{0.9997} & \multicolumn{1}{|c|}{1.00095} \\
		\hline
		\multicolumn{1}{|c}{PARR(p=0.0001)} & \multicolumn{1}{|c}{0.7532} & \multicolumn{1}{|c}{0.9999} & \multicolumn{1}{|c|}{1.00009} \\
		\hline
		\multicolumn{1}{|c}{Lazy correction} & \multicolumn{1}{|c}{0.19$\rightarrow$0} & \multicolumn{1}{|c}{0.362782} & \multicolumn{1}{|c|}{2.177345} \\
		\hline
		\multicolumn{1}{|c}{ADAM} & \multicolumn{1}{|c}{0.5341} & \multicolumn{1}{|c}{0.9807} & \multicolumn{1}{|c|}{1.1765} \\
		\hline
		\multicolumn{1}{|c}{SIWC-size} & \multicolumn{1}{|c}{0.7276} & \multicolumn{1}{|c}{1.0417} & \multicolumn{1}{|c|}{0.9467} \\
		\hline
		\multicolumn{1}{|c}{SIWC-entry} & \multicolumn{1}{|c}{0.0885} & \multicolumn{1}{|c}{1.0628} & \multicolumn{1}{|c|}{0.8951} \\
		\hline
		\multicolumn{1}{|c}{IMDB(e256b8g4)} & \multicolumn{1}{|c}{2.08E-3} & \multicolumn{1}{|c}{0.9561} & \multicolumn{1}{|c|}{0.9937} \\
		\hline
		\multicolumn{1}{|c}{IMDB(e256b8g8)} & \multicolumn{1}{|c}{4.39E-4} & \multicolumn{1}{|c}{0.9560} & \multicolumn{1}{|c|}{0.9941} \\
		\hline
		\multicolumn{1}{|c}{
\begin{small}IMDB(e256b8g4)+FnW
\end{small}} & \multicolumn{1}{|c}{1.66E-3} & \multicolumn{1}{|c}{0.9560} & \multicolumn{1}{|c|}{0.9975} \\
		\hline
		\multicolumn{1}{|c}{\begin{small}IMDB(e256b8g8)+FnW
\end{small}} & \multicolumn{1}{|c}{1.97E-4} & \multicolumn{1}{|c}{0.9560} & \multicolumn{1}{|c|}{0.9977} \\
		\hline
	\end{tabular}
\end{table}

\subsubsection{Write Disturbance Errors} \label{S6.7.1}
The second column in Table~\ref{tab-avg} reports normalized WDEs. PARR shows lower WDEs as the probability scales down, except for p=0.0001. Since rewrite commands might be unnecessary on the infrequently accessed row, excessive restoration with high probability may incur more WDEs (i.e., 41.915 on p=0.1). The lowest probability of 0.0001 in Table~\ref{tab-avg} also leads to more WDEs, because restoration on vulnerable cells is scarce. \textit{Lazy correction} yields non-zero normalized WDE values for different ECPs; however, it is noteworthy that \textit{lazy correction} shows \textit{temporal} WDEs in runtime, which can finally be corrected with ECPs. \textit{SIWC-entry} presents 87.84\% lower WDEs than \textit{SIWC-size} (i.e., 0.0885 vs. 0.7276) because the mitigation performance strongly depends on the cache size. \textit{ADAM} is effective only if the compression ratio exceeds 0.5; hence, \textit{ADAM} shows inferior performance. 

In contrast, IMDB(e256b8g8) reduces WDEs to 4.39E-4, which is \textbf{1218$\times$} and \textbf{202$\times$} fewer WDEs compared to \textit{ADAM} and \textit{SIWC-entry}, respectively. It is noteworthy that these configurations show comparable WDE mitigation performance to the case where the main table consists of 2048 entries without barrier buffers. While a 2048-entry main table requires 108b$\times$2048$\times$4-bank=864KB of SRAM, the combinational approach yields fewer WDEs with a 16KB SRAM, which is four times smaller than \textit{SIWC}. Furthermore, applying FnW to IMDB(e256b8g8) yields 2.2$\times$ fewer WDEs, due to a reduction in the number of bit flips.

\subsubsection{Speedup} \label{S6.7.2}
The third column in Table~\ref{tab-avg} presents the speedup compared to the baseline. PARR achieves similar performance with the baseline regardless of the restoration probability. \textit{Lazy correction} shows the lowest speedup. This is because even though the VnC for corrupted data is deferred, at least four read operations strictly ordered by a write command are necessary. Although the proposed method rewrites two neighbors, these operations are performed in an on-demand fashion instead of incurring four read operations per write operation, as VnC does. Therefore, the proposed method can outperform \textit{lazy correction}. The speed of \textit{ADAM} degrades by about 2\% due to encoding and decoding processes of FPC. For \textit{SIWC-entry} and \textit{-size}, slightly higher performance is achieved.  

On the other hand, two configurations of the proposed method experience approximately 4\% speed degradation on average. The waiting cycles for memory systems constitute 12\% of execution time in the baseline, according to our evaluation. Consequently, the proposed method degrades the performance of the overall system only by \textbf{0.48\%}. 
IMDB requires 1-3 cycles for processing a write command on the critical path. The latency is determined by the hit/miss cases of the main table and the barrier buffer. If a write command hits on the main table, one referring cycle is spent on the main table. Furthermore, suppose this hit command triggers the rewrite operation. In that case, one more read cycle on the barrier buffer is required, because the hit entry must be promoted to the barrier buffer (i.e., contents swapping). Finally, the swapped contents are written to the main table and the barrier buffer, maximally resulting in three cycles. If a write command misses, AppLE must be performed for finding a replacing candidate. However, the latency of AppLE can be hidden within the write latency (i.e., IDLE state in Figure~\ref{fig-AppLE} (c)). In a memory system, all commands follow promised timing constraints (i.e., JEDEC DDR standards). Thus, the media controller must wait for the latency (i.e., 150 ns or 120 cycles) after issuing a write command to a bank. As a result, the negligible latency of IMDB leads to minor performance degradation.

\subsubsection{Energy} \label{S6.7.3}
The fourth column in Table~\ref{tab-avg} shows the normalized energy. PARRs show higher energy consumption than the baseline for all probabilities, because rewrite commands cause higher write energy consumption. However, the energy overhead is not notable, due to relatively low probabilities (i.e., $\leq$0.1) of PARRs. \textit{Lazy correction} consumes 2.18$\times$ higher energy than the baseline, because both execution time and the number of commands increase. Meanwhile, \textit{SIWC-size} reduces 5\% of energy compared to the baseline. This is because persistent workloads have relatively high locality due to cache line flush instructions, reducing write operations on frequently accessed addresses. Furthermore, the energy can be reduced by about 10.5\% compared to the baseline with a larger number of entries, as declared by \textit{SIWC-entry}; however, it should be noted that the WDE mitigation performance is not as excellent as it is with the proposed methods. 
Although IMDB(e256b8g8) presents 9\% higher energy consumption compared to \textit{SIWC-entry}, this outcome is still \textbf{0.59\%} smaller than the baseline. Even though the proposed scheme generates rewrite commands that may contribute to the energy consumption, the “tiny” barrier buffer reduces the write traffic with a 10.67\% cache hit rate, leading to lower energy consumption than the baseline. In contrast, IMDB(e256b8g8) consumes \textbf{54.4\%} less energy than \textit{lazy correction}. 

\section{Discussion} \label{S7}
\begin{figure} [t]
\centering
\includegraphics[scale=0.75]{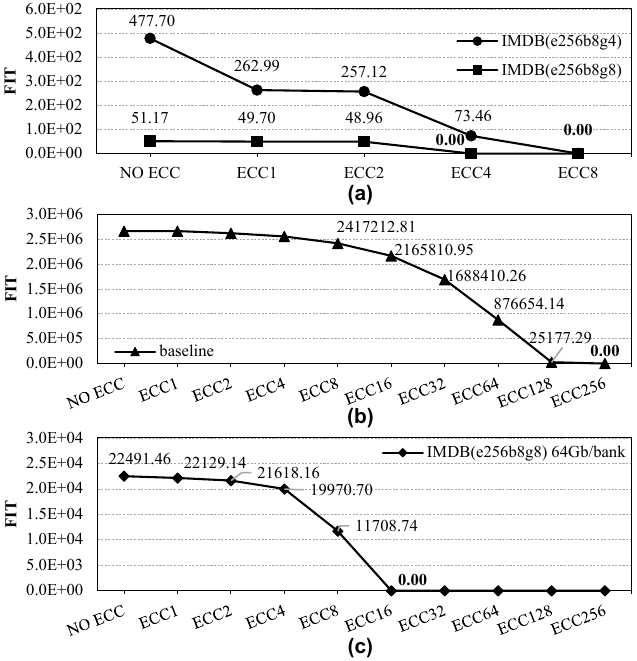}
\caption{FITs when different ECC schemes are applied to (a) IMDB, (b) baseline, (c) IMDB(e256b8g8) for a 64Gb bank.}
\label{fig-ECC}
\end{figure}

\textbf{Synergy with ECC schemes}.
In general, error-correcting codes (ECC) are proactively being employed in memory products that have reliability-related problems. In our case, ECC logic is placed on the media controller for system expandability. To observe the system reliability, we evaluated failure-in-time (FIT), which is the number of corrupted bits in an hour \cite{FormulaFIT, IntroFIT}. Commonly, Figure~\ref{fig-ECC} shows that FITs decrease when the correction capability of ECC enhances. In particular, Figure~\ref{fig-ECC} (a) shows that 0-FIT can be achieved when ECC4 (i.e., 4-bit error correction) and ECC8 (i.e., 8-bit error correction) are applied to IMDB (e256b8g8) and IMDB (e256b8g4), respectively. A (552, 512)-BCH code that is capable of correcting 4 errors \cite{NVMCircuitLevelECC} only incurs 1.5ns of latency (i.e., $<$1 cycle at 800MHz), according to the latency formula in \cite{BCHForMem}. Therefore, only a minuscule amount of latency is required when IMDB is assisted by ECC. 
Figure~\ref{fig-ECC} (b) shows that ECC-16 is ineffective for WDEs. We observe that simultaneous bit flips occur in one data across all workloads, leading to simultaneous WDEs in multiple cells. Since ECC has no knowledge of such programming patterns, ECC is incapacitated by WDEs.
In contrast, Figure~\ref{fig-ECC} (b) shows that ECC256 yields 0-FIT. The correction capability of ECC256 corresponds to a (3584, 512)-BCH code. This code yields 611$\times$ larger area than that of (552, 512)-BCH code according to the area formula in \cite{BCHForMem}. Therefore, IMDB is necessary for obtaining a reliable memory system with a lower area burden on ECC.

\textbf{Discussion of SRAM capacity against SIWC}. 
Considering the capacity of SRAM for the proposed method and a write cache-based study (i.e., \textit{SIWC}) in a four-bank PCM system, the latter requires 256$\times$64B$\times$4-bank=64KB of SRAM if 256 addresses are managed per bank. On the other hand, for the proposed method, the main table entry has 25b+8b+72b +3b=108b, and the barrier buffer entry has 64B+25b+8b+8b =553b (see Figure~\ref{fig-IMDB} (a)). Therefore, the proposed method requires 256$\times$108b$\approx$3.4KB of SRAM on the main table per PCM bank. 
We evaluate our system by configuring the main table as a fully associative SRAM, because a 3.4KB of fully associative SRAM cache does not burden resources. In addition, a fully associative can yield the best performance compared to fewer ways. 
The barrier buffer consumes 8$\times$553b$\approx$0.6KB of SRAM per PCM bank (see Section~\ref{S6.5}). Consequently, (3.4KB+0.6KB)$\times$4-bank=16KB of SRAM translates to 2KB per 1GB of PCM. If 256 addresses are managed, the proposed method consumes \textbf{4$\times$} smaller SRAM area than \textit{SIWC}, and the gap enlarges as the number of managed addresses grows. Besides the SRAM capacity, introducing SRAM as a data region requires considering the hold-up time constraint of supercapacitors. In particular, \textit{SIWC} only holds dirty data; hence, flushing 256 volatile data requires 150ns$\times$256 flushes/ 100us=38.4\% of flush time at most (i.e., all row buffer miss commands on a single bank), where the value of 100us comes from \cite{NVSLOptane}. In contrast, flushing data in the barrier buffer only requires 150ns$\times$8 flushes/100us=1.2\%. In conclusion, IMDB mitigates more WDEs without expanding supercapacitors.

\textbf{Discussion of area overhead}.
\rev{IMDB mainly consists of a main table, a barrier buffer, control logic for AppLE, and control logic for integrated counter blocks. First, the main table and the barrier buffer require 16 KB of SRAM, which translates to 768K transistors considering 6T SRAM. Second, the control logic for AppLE requires a 9-bit comparator. The comparator consists of one AND-gate, one NOR-gate, and 26 AND-gates with one bubbled input \cite{9bComparator}, each requiring 6, 4, and 10 transistors, respectively. Thus, the 9-bit comparator consists of a total of 270 transistors (=6+4+26$\times$10). Lastly, the control logic for integrated counter blocks in Figure 3 (b) consists of 5.5M transistors according to synthesis results from Synopsys Design Compiler. Consequently, IMDB consists of 6.268M transistors in total. We find that a representative DRAM controller in \cite{PARDIS} requires approximately 3.7B transistors (i.e., 1.8 mm$^2$ at 22 nm). Therefore, IMDB incurs 0.17\% of area overhead with respect to the representative DRAM controller. It is noteworthy that the PCM controller area is not disclosed; however, the higher complexity of the PCM controller than that of DRAM explicitly proves that IMDB occupies a small amount of area. }

\textbf{Scalability of the proposed scheme}.
We evaluate WDEs for a larger bank density compared to a 2GB (i.e., 16Gb) bank that is adopted in Section~\ref{S6} to observe the scalability of IMDB. The normalized WDE of a 64Gb bank with IMDB (e256b8g8) is 9.27E-3, 20$\times$ higher than a 16Gb bank (i.e., 4.38E-4). This result indicates that the currently proposed size has less effect on a larger density, because an IMDB plane should manage more addresses. We can address such a scalability issue in two ways. First, enlarging the number of entries in the main table to 512 achieves 4.54E-4 WDEs, which is again similar to the result of the original 16Gb bank with IMDB(e256b8g8) (i.e., 4.38E-4). Second, stronger ECC can be applied to mitigate WDE in a larger density bank. 
Figure~\ref{fig-ECC} (c) presents FITs when various ECC schemes are applied to the 64 Gb bank PCM, which is supported by IMDB(e256b8g8). For achieving 0-FIT in a 64Gb PCM, ECC16 is necessary rather than ECC4, which is effective in the 16 Gb bank PCM (see Figure~\ref{fig-ECC} (a)). 

\section{Related Work} \label{S8}
\textbf{VnC-based schemes}. VnC is the most solid method capable of preventing WDEs \cite{VnCinSTT, SDPCM}; it triggers two pre-write read operations and two post-write read operations, before and after a write operation, respectively. These four read commands are strictly ordered by one write command, incurring significant performance overhead. In \cite{SDPCM}, \textit{lazy correction} temporarily stores the locations of disturbed cells in an error-correction pointer (ECP) chip, deferring the subsequent VnC as late as possible until the ECP becomes full. However, cells in the ECP must be well insulated to guarantee no errors. Also, it is necessary to execute at least four read operations for the initial write command.

\textbf{Encoding-based schemes}. Data encoding can reduce WDE-vulnerable patterns \cite{DIN, MinWD, ADAM, WLC, CEnT, ADAPT}. In \cite{DIN}, \textit{DIN} proposes a codebook that encodes contiguous 0s in a compressed pattern to eliminate patterns vulnerable to WDEs. However, this approach must fall back on the VnC method if the length of the encoded data exceeds the length of the cache line. In \cite{MinWD}, \textit{MinWD} encodes write data into multiple candidates with special shift operations and selects the least aggressive form from all candidates. However, this method requires additional bits as an indicator of the shift operation. In \cite{ADAM}, \textit{ADAM} compresses a cache line and aligns the line to the right and left alternately; hence, the number of valid bits on adjacent rows is reduced. However, encoding schemes strongly depend on the data patterns of the applications. \textit{WLC} \cite{WLC} is a compression scheme for reducing energy; it compresses few MSBs of each 64-bit of a cache line, increasing "the number lines" to be compressed. However, compared with the compression ratio of 40\% in \textit{ADAM}, the compression ratio of \textit{WLC} is bounded to 9$\times$8/512=14.1\% if 9 MSBs in each 64-bit can be compressed. Thus, WLC is less effective than ADAM. 

\textbf{Cache-based scheme}. Storing frequently updated data in volatile caches can enhance the system reliability. In \cite{SIWC}, \textit{SIWC} leverages a write cache that inserts data probabilistically and absorbs bit flips. Because WDE-vulnerable data would be stored in the write cache, the victims of WDEs become safe. However, this method introduces several mega-bytes of volatile memory to obtain a high hit ratio, and the supercapacitor for data flushes must be expanded. Furthermore, \textit{SIWC} reports the number of operations that may incur WDEs (i.e., WDE limitation number), but this information is not utilized for WDE mitigation.

\section{Conclusion} \label{S9}
WDE is a severe reliability problem that hinders the manufacturing of PCMs. This study proposes a table-based approach, IMDB, to restore cells on demand within a module. The newly proposed replacement policy yields higher reliability than the LRU and fully randomized replacement policies. Subsequently, AppLE enables an efficient implementation of the replacement policy. The small barrier buffer absorbs bit flips, offloading the burden onto the supercapacitor. Consequently, rigorous sensitivity analyses concerning design parameters are conducted to obtain a cost-effective architecture. The evaluation results show that the proposed method significantly reduces WDEs compared to the outcomes of earlier studies while maintaining speed and energy consumption levels that approximate those of the baseline. 

\section*{ACKNOWLEDGMENTS}
We thank the anonymous reviewers to substantially improve the paper. Also, special appreciations go to Jiwoong Choi, Boyeal Kim, and Taehyun Kim for their feedback.
This work was partly supported by the National Research Foundation of Korea(NRF) grant funded by the Korea government(MSIT) (No. 2022R1F1A1062786), the MSIT (Ministry of Science and ICT), Korea, under the ITRC (Information Technology Research Center) support program (IITP-2022-2020-0-01461) supervised by the IITP(Institute for Information \& communications Technology Planning \& Evaluation), and the Technology Innovation Program (20011074, Development of Open Convergence Memory Solution and Platform for Next Generation Memories) funded by the Ministry of Trade, Industry \& Energy(MOTIE, Korea)
The EDA tool was supported by the IC Design Education Center (IDEC), Korea.

%%%%%%% -- PAPER CONTENT ENDS -- %%%%%%%%

%%%%%%%%% -- BIB STYLE AND FILE -- %%%%%%%%
%\newlength{\bibitemsep}\setlength{\bibitemsep}{0pt}
%\newlength{\bibparskip}\setlength{\bibparskip}{0pt}
%\let\oldthebibliography\thebibliography
%\renewcommand\thebibliography[1]{%
%  \oldthebibliography{#1}%
%  \setlength{\parskip}{\bibitemsep}%
%  \setlength{\itemsep}{\bibparskip}%
%}
\bibliographystyle{IEEEtranS}
\bibliography{refs}

\end{document}